\documentclass[preprint]{cimento}

\usepackage{graphicx}  
\usepackage{amsmath,amssymb}

\newcommand{\Dslash}{\!\not\!\! D}

\title{New Physics at the LHC: Strong vs Weak symmetry breaking}
\author{R.~Contino\from{ins:x}}
\instlist{\inst{ins:x} Dipartimento di Fisica, Universit\`a di Roma ``La Sapienza'' and \\
INFN Sezione di Roma, I-00185 Roma, Italy}
\begin{document}

\maketitle

\begin{abstract}
What kind of New Physics, if any, we expect to discover at the LHC?
I will try to address this formidable question by re-formulating it as
follows: is the breaking of the electroweak symmetry strong or weak ?
\end{abstract}

\section{The need for an Electroweak Symmetry Breaking sector}

The entire physics discovered so far in high-energy experiments can be
described and compactly summarized by the Lagrangian 
\begin{equation}
\label{eq:sofar}
\begin{split}
& {\cal L} = {\cal L}_0 + {\cal L}_{mass} \\[0.2cm]
& {\cal L}_0 = -\frac{1}{4} W^a_{\mu\nu} W^{a\, \mu\nu} -\frac{1}{4} B_{\mu\nu} B^{\mu\nu} - 
   \frac{1}{4} G_{\mu\nu} G^{\mu\nu} + \sum_{j=1}^3 \left( 
   \bar\Psi_L^{(j)} i\Dslash \Psi_L^{(j)}  + \bar\Psi_R^{(j)} i\Dslash \Psi_R^{(j)} \right) \\
& {\cal L}_{mass} = M_W^2\, W^+_{\mu} W^{-\, \mu} + \frac{1}{2} M_Z^2\, Z^\mu Z_\mu \\
   & \qquad\qquad - \sum_{i,j} \left( \bar u^{(i)}_L M^u_{ij} u_R^{(j)} + \bar d^{(i)}_L M^d_{ij} d_R^{(j)}
   + \bar e^{(i)}_L M^e_{ij} e_R^{(j)} + \bar \nu^{(i)}_L M^\nu_{ij} \nu_R^{(j)} \right) + h.c. \, ,
\end{split}
\end{equation}
where $\Psi$ is a collective index for the Standard Model (SM) fermions
and $i,j$ are generation indices.
A remarkable property that has emerged, and that is most clearly exhibited by the Lagrangian ${\cal L}$,
is that while all the fundamental interactions among the particles are symmetric under a local
$SU(2)_L\times U(1)_Y$ invariance, the observed mass spectrum is not.
In other words,  the electroweak symmetry is hidden,  i.e. spontaneously broken by the vacuum.
In mathematical terms, the spontaneous breaking can be made more explicit by introducing 
as propagating degrees of freedom the Nambu-Goldstone bosons $\chi^a$ that 
correspond to the longitudinal polarizations of the $W$ and $Z$ bosons (for simplicity, from here on
I will omit the lepton terms and concentrate on the quark sector):
\begin{equation}
\label{eq:chiralLmass}
\begin{split}
&\Sigma(x) = \exp(i\sigma^a \chi^a(x)/v), \, \qquad D_\mu \Sigma = \partial_\mu \Sigma 
 -i g_2 \frac{\sigma^a}{2} W^a_\mu \Sigma + i g_1 \Sigma \frac{\sigma^3}{2} B_\mu \\[0.3cm]
& {\cal L}_{mass} = \frac{v^2}{4} \text{Tr}\left[ \left( D_\mu \Sigma \right)^\dagger
 \left( D_\mu \Sigma \right) \right] - \frac{v}{\sqrt{2}} \sum_{i,j} \left( \bar u_L^{(i)} d_L^{(i)} \right) 
 \Sigma \begin{pmatrix} \lambda_{ij}^u\,  u_R^{(j)} \\[0.1cm] \lambda_{ij}^d\,  d_R^{(j)} \end{pmatrix} + h.c.
\end{split}
\end{equation}
The local $SU(2)_L\times U(1)_Y$ invariance is now manifest in the Lagrangian (\ref{eq:chiralLmass}) 
with $\Sigma$ transforming as 
\begin{equation}
\Sigma \to U_L(x) \, \Sigma\, U_Y^\dagger(x)\, , \qquad U_L(x) = \exp \big(i\alpha_L^a(x) \sigma^a/2 \big)
\quad   U_Y(x) = \exp \big(i\alpha_Y(x) \sigma^3/2 \big)\, .
\end{equation}
In the unitary gauge $\langle \Sigma \rangle = 1$, the chiral Lagrangian (\ref{eq:chiralLmass}) reproduces
the mass term of eq.(\ref{eq:sofar}) with
\begin{equation}
\label{eq:rho}
\rho \equiv \frac{M_W^2}{M_Z^2 \cos^2\theta_W} = 1 \, ,
\end{equation}
which is consistent with the experimental measured value to quite good accuracy.
The above relation follows as the consequence of a larger global $SU(2)_L\times SU(2)_R$
approximate invariance of (\ref{eq:chiralLmass}), $\Sigma \to U_L \, \Sigma\, U_R^\dagger $,
which is spontaneously broken to the diagonal subgroup  $SU(2)_c$ by $\langle \Sigma \rangle = 1$, and
explicitly broken by $g_1$ and $\lambda^u_{ij} \not = \lambda^d_{ij}$.
In the limit of vanishing $g_1$ the ``custodial'' $SU(2)_c$ implies $M_W = M_Z$, which is
replaced by eq.(\ref{eq:rho}) at tree level for arbitrary $g_1$.
Further corrections proportional to $g_1$ and $\lambda^u - \lambda^d$ arise at the one-loop level 
and are small. In fact, the success of the tree-level prediction $\rho =1$  a posteriori justifies 
the omission in the chiral Lagrangian~(\ref{eq:chiralLmass}) of the additional term 
\begin{equation}
 c_3\, v^2 \, \text{Tr}\left[ \Sigma^\dagger D_\mu \Sigma \, T^3 \right]^2 
\end{equation}
that is invariant under 
the local $SU(2)_L\times U(1)_Y$ but explicitly breaks the global $SU(2)_L\times SU(2)_R$.
In other terms, the coefficient $c_3$ of such extra operator
is experimentally constrained to be very small.

The chiral formulation (\ref{eq:chiralLmass}) makes  the limit of our current 
mathematical description most transparent: There is a  violation of  perturbative unitarity in the scattering 
$\chi\chi \to \chi\chi$ at energies $E\gg M_W$, which is
ultimately linked to the non-renormalizability of the chiral Lagrangian.
More specifically, the scattering amplitude grows with $E^2$,
\begin{equation}
\begin{gathered}
A(\chi^a \chi^b \to \chi^c \chi^d) = 
 A(s)\, \delta^{ab}\delta^{cd} +  A(t)\,\delta^{ac}\delta^{bd} + A(u)\, \delta^{ad}\delta^{bc}\, , \\[0.3cm]
 A(s) = \frac{s}{v^2} \left[ 1 + O\left(\frac{M_W^2}{s}\right) \right]
\end{gathered}
\end{equation}
due to the derivative interaction among four Goldstones that comes
from expanding the  kinetic term of $\Sigma$ in eq.(\ref{eq:chiralLmass}).
Intuitively, the
$\chi$'s are the degrees of freedom that are eaten in the unitary gauge to form
the longitudinal polarizations of $W$ and $Z$. The scattering of four Goldstones
thus corresponds
to the physical scattering  of four longitudinal vector bosons:  $V_L V_L \to V_L V_L$, with  $V_L = W_L, Z_L$.
Such correspondence is made formal by the
Equivalence Theorem, which states that the amplitude for the emission or absorption of a Goldstone field $\chi$
becomes equal at large energies to the amplitude for the emission or absorption of a longitudinally-polarized vector 
boson.
As a consequence, the physical scattering $V_L V_L \to V_L V_L$ violates perturbative unitarity at large energies 
$E\gg M_W$, and the leading energy behavior of its cross section is captured by that of the easier process 
$\chi\chi \to \chi\chi$. The merit of the chiral formulation is that of 
isolating the problem to the sector of the 
Lagrangian which leads to the 
mass terms for the vector bosons and the fermions. In other terms:

\vspace{0.3cm}
\textit{We need some new  symmetry breaking dynamics that acts as an ultraviolet 
completion of the electroweak chiral Lagrangian~(\ref{eq:chiralLmass}) and restores unitarity.}
\vspace{0.3cm}

As required by the experimental evidence, such new dynamics must be (approximately) custodially symmetric,
so as to prevent large corrections to the $\rho$ parameter.
The question then arises: is this dynamics weak or strong  ?

\section{Strong vs Weak symmetry breaking}

The most economical example of electroweak symmetry breaking (EWSB) sector is that of just one new scalar field $h(x)$,
singlet under  $SU(2)_L\times SU(2)_R$,
in addition to the Goldstones $\chi$. Assuming that $h$ is coupled to the SM gauge fields and
fermions only via weak gauging and (proto)-Yukawa couplings, the most general EWSB Lagrangian
has three free parameters $a$, $b$, $c$~\footnote{In general $c$
will be a matrix in flavor space, but in the following we will assume for simplicitly 
just one fermion generation. The extension to three families is 
straightforward.} at the quadratic order in $h$~\cite{stronghh}
\begin{equation}
\label{eq:CHLag}
\begin{split}
{\cal L}_{EWSB} =& \frac{1}{2} \left(\partial_\mu h\right)^2 + V(h) + 
 \frac{v^2}{4} \text{Tr}\left[ \left( D_\mu \Sigma \right)^\dagger \left( D_\mu \Sigma \right) \right] 
 \left( 1 + 2 a\, \frac{h}{v} + b\, \frac{h^2}{v^2} + \dots \right) \\
 & - m_\psi \, \bar\psi_L \Sigma \left( 1+ c\, \frac{h}{v} + \cdots\right) \psi_R + h.c.
\end{split}
\end{equation}
where $V(h)$ is some potential, including a mass term, for $h$.
Each of these parameters controls the unitarization of a different sector of the theory:
For $a=1$ the exchange of the scalar unitarizes the $\chi\chi\to\chi\chi$ ($V_L V_L\to V_L V_L$) scattering 
(in the following, dashed and solid lines denote respectively the fields $\chi$ and $h$, whereas 
solid lines with an arrow denote fermions) 
\\[0.5cm]
\hspace*{0.5cm}
\includegraphics[height=17mm]{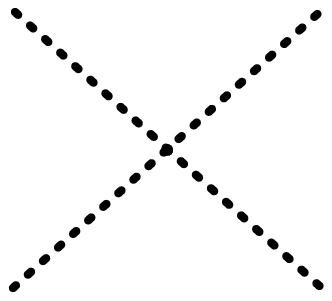} \hspace{1cm}
\includegraphics[height=17mm]{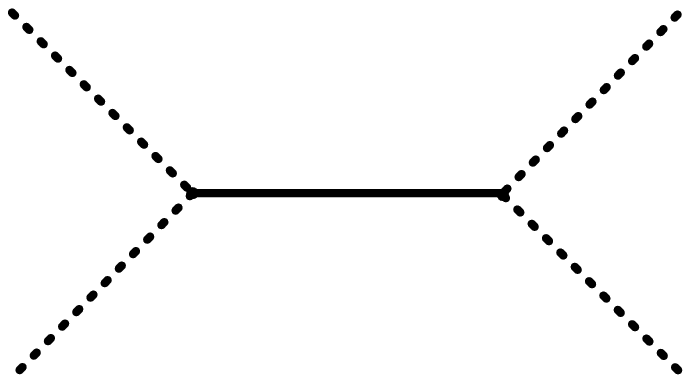} \hspace{1cm}
\begin{minipage}{0.1\linewidth}
  \vspace*{-2cm}
  \begin{equation*}
  A(s) \simeq \frac{s}{v^2} \left(1-a^2\right)\, ; 
  \end{equation*}
\end{minipage}  
\\[0.4cm]
For $b=a^2$ also the inelastic channel $\chi\chi\to hh$ ($V_L V_L\to hh$) respects unitarity \\
\hspace*{0.5cm}
\includegraphics[height=17mm]{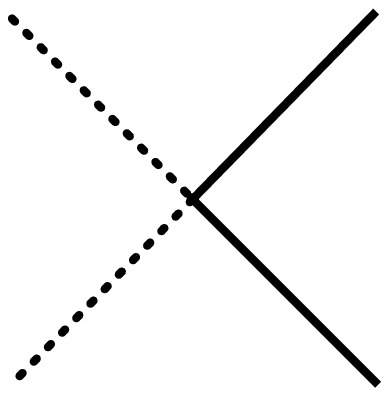} \hspace{1cm}
\begin{minipage}{0.15\linewidth}
  \vspace*{-1.5cm}
  \includegraphics[height=23mm]{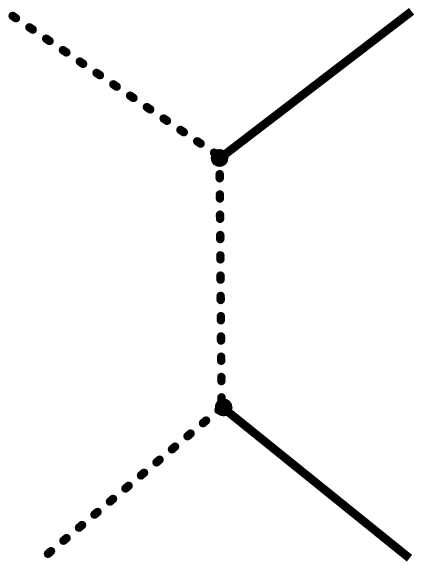} 
\end{minipage} \hspace{2.3cm}
\begin{minipage}[h]{0.2\linewidth}
\vspace*{-1.9cm}
\begin{equation*}
A(\chi^a\chi^b\to hh) \simeq \delta^{ab} \,\frac{s}{v^2} \left(b-a^2\right) \, ;
\end{equation*} 
\end{minipage} \\[0.4cm]
Finally, for $ac=1$ the $\chi\chi\to\psi\bar\psi$ ($V_L V_L \to\psi\bar\psi$) scattering is unitarized \\[0.5cm]
\hspace*{0.5cm}
\includegraphics[height=17mm]{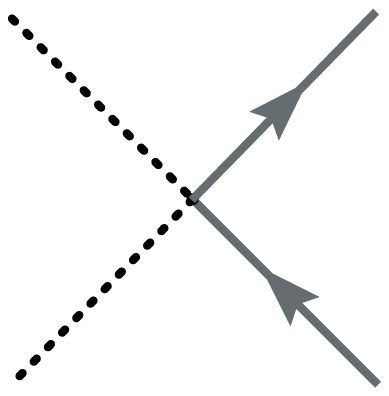} \hspace{1cm}
\includegraphics[height=17mm]{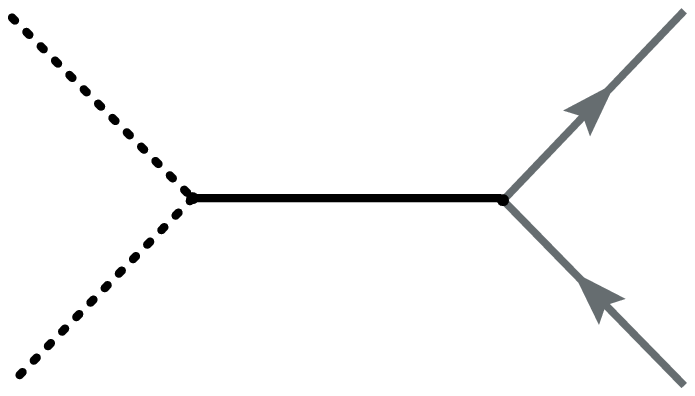} \hspace{1cm}
\begin{minipage}[h]{0.4\linewidth}
\centering
\vspace*{-2cm}
\begin{equation*}
A(\chi^a\chi^b\to \psi\bar\psi) \simeq \delta^{ab} \,\frac{\sqrt{m_\psi s}}{v^2} \left(1-a c \right) \, .
\end{equation*} 
\end{minipage}
\\[0.4cm]
Only for $a=b=c=1$ 
the EWSB sector is  weakly interacting (provided the scalar $h$ is light), 
as for example $a\not =1$ implies 
a strong $VV\to VV$ scattering with violation of perturbative unitarity at energies 
$\sqrt{s} \approx 4\pi v/\sqrt{1-a^2}$, and similarly for the other channels.
The point $a=b=c=1$ in fact defines what I would call the ``Higgs model'':
${\cal L}_{EWSB}$ (with vanishing higher-order terms in $h$) 
can be rewritten in terms of the $SU(2)_L$ doublet
\begin{equation}
H(x) = \frac{1}{\sqrt{2}}\, e^{i \sigma^a \chi^a(x)/v} \begin{pmatrix} 0 \\ v + h(x) \end{pmatrix}
\end{equation}
and gets the usual form of the  Standard Model Higgs Lagrangian.
In other words, $\chi^a$ and $h$ together form a linear representation of  $SU(2)_L\times SU(2)_R$.
In terms of the Higgs doublet $H(x)$, the custodial invariance of the Lagrangian
appears like an accidental symmetry:  at the renormalizable
level, all the  ($SU(2)_L\times U(1)_Y$)-invariant operators are functions of $H^\dagger H = \sum_i \omega_i^2$,
where $\omega_i$ are the four real components parametrizing the complex doublet $H$.
An $SO(4)\sim SU(2)_L\times SU(2)_R$ invariance follows under which these components are rotated,  broken to 
$SO(3)\sim SU(2)_c$ in the vacuum $\langle H^\dagger H\rangle = v^2$.
The unitarity of the Higgs model can be traced back to its renormalizability, which is now evident 
from the Lagrangian written in terms of $H$.

The weakly-interacting Higgs model has two main virtues: it is theoretically attractive because of its
calculability, and it is insofar phenomenologically successful, passing in particular all the LEP and SLD 
electroweak precision tests.
Both calculability  (which stems from  perturbativity) and the success in passing the precision tests follow
from the Higgs boson $h$ being light. 
It is however well known that an elementary light scalar, such as $h$, is  unstable under
radiative corrections, hence highly unnatural in absence of some symmetry protection.
It is quite possible, on the other hand, that a light Higgs-like scalar arises 
as a bound state of a new strong dynamics:
its being composite would solve the SM Higgs hierarchy problem, while its being light would
still be required to pass the electroweak tests.
The Lagrangian (\ref{eq:CHLag}) with generic $a$, $b$, $c$ gives a general parametrization of such 
composite Higgs theories where all the other resonances have been integrated out.
Away from the unitary point $a=b=c=1$ the exchange of the light composite Higgs $h$ 
fails to completely
unitarize the theory, which eventually becomes strongly interacting at high energies.
Similarly to pion-pion scattering  in QCD,
unitarity is ultimately reinforced at the strong dynamics scale through the exchange of 
the other (spin-1) resonances.

\begin{figure}
\begin{center}
\begin{minipage}[b]{3.5cm}
\includegraphics[height=27mm]{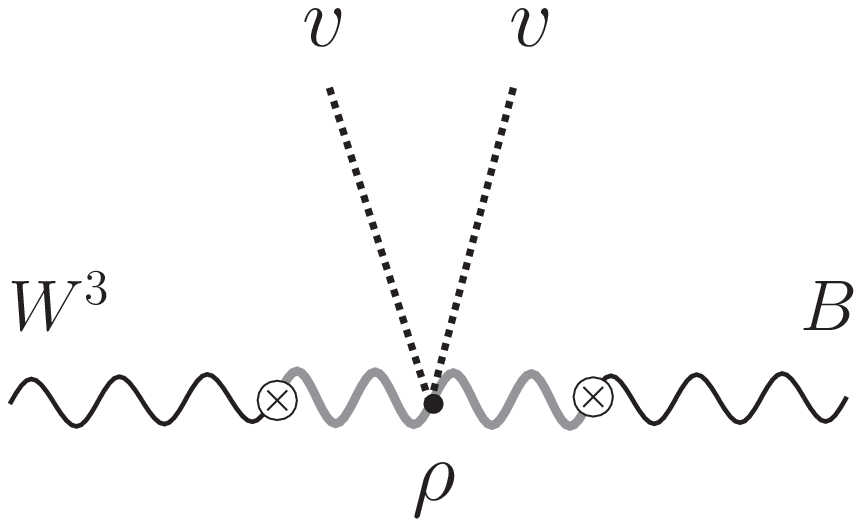} \\[1cm]
\includegraphics[height=17mm]{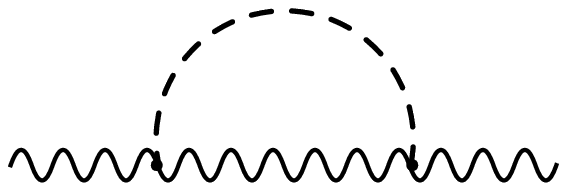} 
\end{minipage}
\hspace{2.0cm}
\begin{minipage}[b]{6.5cm}
\includegraphics[height=60mm]{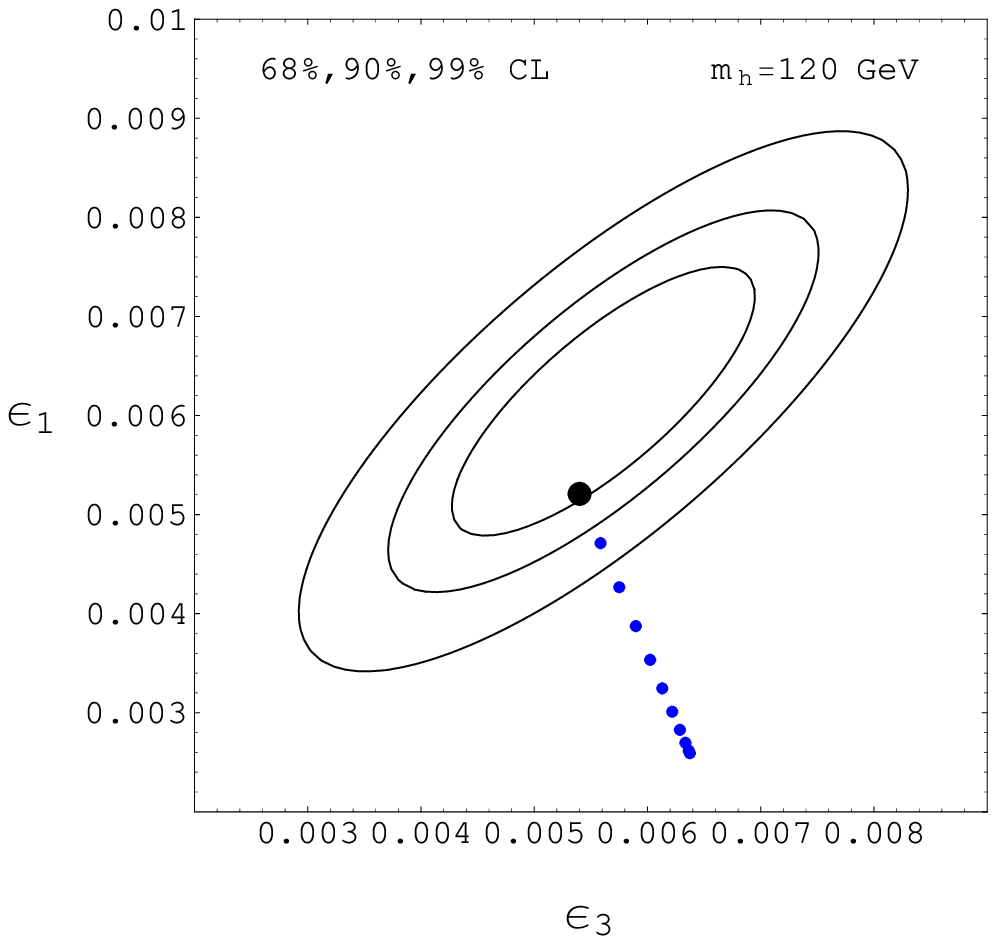}
\vspace*{-0.5cm}
\end{minipage}
\end{center}
\caption{\label{fig:Spar}
Upper left: Contribution to the $S$ parameter from the exchange of heavy spin-1 resonances.
Lower left: One-loop Higgs contribution to $S$ and $T$.
Right: Correction to $\epsilon_{1,3}$ 
as a consequence of the modified Higgs coupling  
to the SM vectors: 
$\Delta \epsilon_{1} = \alpha\, T = - c_1 (1-a^2) \log (\Lambda^2/m_h^2)$,
$\Delta \epsilon_{3} = \alpha/(4 \sin^2\!\theta_W)\, S = - c_3 (1-a^2) \log (\Lambda^2/m_h^2)$.
The black fat dot shows the SM prediction ($a=1$) for $m_h = 120\,$GeV.
The smaller blue dots show how $\epsilon_{1,3}$ are modified by diminishing $a$ from 1 to 0 in steps
of 0.1 (for $\Lambda = 2.5\,$TeV). The $68\%$, $90\%$, $99\%$ CL contours have been obtained by 
setting $\epsilon_2$ and $\epsilon_b$ to their SM value.
}
\end{figure}

Insofar we have tacitly assumed that these latter are heavier than the composite Higgs.
This is in fact required (unless some non-trivial symmetry protection mechanism is at work)
to avoid large corrections to the precision observables, for example to the Peskin-Takeuchi
$S$ parameter, see fig.~\ref{fig:Spar}.
As first pointed out by Georgi and Kaplan in the eighties~\cite{GK}, the composite Higgs
boson can  be naturally lighter than the other resonances if it 
emerges as the (pseudo-)Goldstone boson of an enlarged global symmetry of the strong dynamics.
For example, if the strong sector has an $SO(5)$ global symmetry spontaneously broken down to
$SO(4)\sim SU(2)_L\times SU(2)_R$, this implies four real Goldstones trasforming as a fundamental
of $SO(4)$, or equivalently as a complex doublet $H$ of $SU(2)_L$~\cite{Agashe:2004rs}.
The couplings of the SM fermion and gauge fields to the strong sector will in general
break explicitly  its global symmetry, thus generating a Higgs potential at the one-loop level.
By naive dimensional analysis, the expected mass scale of the other resonances
is $m_\rho \sim 4\pi f$, where $f$ is the $\sigma$-model scale associated with the
composite Higgs. This latter gets instead a much lighter mass $m_h \sim g_{SM} v$ at one-loop,
with $g_{SM}$ being some SM coupling, thus implying a parametric hierarchy $m_h \ll m_\rho$.  
In this context, the electroweak scale $v$ is dynamically determined
and will not in general coincide with~$f$, differently from Technicolor theories
where no separation of scales exists. The ratio  $\xi = v^2/f^2$ sets the size of the parametric 
suppression in all corrections to the precision observables, as $f\to\infty$ ($\xi\to 0$) with fixed $v$
is a decoupling limit where the Higgs stays light and all the other resonances become
infinitely heavy.  As a matter of fact, $v\lesssim 0.3 f$ is enough to largely suppress
any correction from the heavy resonances.

\section{The Higgs boson: elementary or composite?}

It is at this point clear that the discovery of a light Higgs boson alone will not 
be sufficient to rule out the possibility of a strong electroweak symmetry breaking.
Experimentally, one should measure the parameters $a$, $b$, $c$ as precisely as possible
and look for deviations from the unitary point $a=b=c=1$.
In general these parameters are independent from each other, although they will be
related in specific composite Higgs models as functions of $\xi$. Ref.~\cite{Giudice:2007fh} 
also showed that the behavior of $a$ and $b$ at small $\xi$ is universal whenever the light
Higgs boson is part of a composite $SU(2)_L$ doublet. 

A first determination of $a$ and $c$ will come from a precise measurement of the couplings
of the Higgs to the SM fermions and vectors. This will require disentangling possible
modifications of both the Higgs production cross sections and decay rates.
Preliminary studies have shown that the LHC should be eventually able to extract the individual
Higgs couplings with a $\sim 20\%$ precision~\cite{higgscoupl}, though much will depend on the value
of its mass.  This would imply a sensitivity on $(1-a)$ up to $0.1 - 0.2$~\cite{Giudice:2007fh}.
%
%
As stressed by the authors of ref.~\cite{Barbieri:2007bh}, the parameter $a$ is already 
constrained by the LEP precision data:
modifying the Higgs coupling to the SM vectors changes the infrared
one-loop contribution to $\epsilon_{1,3}$ ($\epsilon_1 = \epsilon_1^{SM} + \alpha\, T$,
$\epsilon_3 = \epsilon_3^{SM} + \alpha/(4 \sin^2\!\theta_W)\, S$) by an amount 
$\Delta \epsilon_{1,3} =- c_{1,3} (1-a^2) \log (\Lambda^2/m_h^2)$, 
where $c_{1,3}$ are one-loop numerical coefficients and $\Lambda$ denotes
the scale at which the other resonances set on and unitarity is ultimately restored in $VV$ scattering.
For example, assuming no additional corrections to the precision observables and setting
$m_h = 120\,$GeV, $\Lambda = 2.5\,$TeV, one obtains $a \gtrsim 0.85$ at 99$\%$~CL, see fig.~\ref{fig:Spar}. 

Measuring the Higgs couplings will give important clues on its nature
and on the role it plays in the EWSB mechanism. A ``direct'' probe of the strenght of the symmetry breaking
dynamics will however come only from a precise study of the $VV$ scattering.
A smoking gun of strong electroweak symmetry breaking would be discovering a light Higgs and 
at the same time finding
an excess of events in $VV\to VV$ scattering compared to the SM expectation: this would be
the sign that the energy-growing behavior of the scattering cross section of longitudinal $W$ and $Z$'s
is not saturated at a low scale by the Higgs exchange. As a matter of fact, it turns out 
that the onset of the strong scattering is delayed to higher energies due to a large
``pollution'' from the scattering of transverse polarizations, (see for example~\cite{stronghh}).
This requires defining the signal as 
the difference between the observed number of events and the
one predicted in the SM
for a reference value of the Higgs mass~\cite{Bagger:1993zf}, since the contribution
from the transverse scattering is not expected to  change significantly in presence of new physics:  
$\sigma( V_LV_L\, \text{signal}) = \sigma(VV\, \text{observed}) - \sigma(SM)$.
The pioneering preliminary study of ref.~\cite{Bagger:1993zf} 
focussed on the ``golden'' purely leptonic decay modes and
showed that perhaps the best channel to observe a strong scattering
signal is $W^{\pm} W^{\pm} \to W^{\pm} W^{\pm}$ (all $W$'s with the same charge), see table~\ref{tab:WWscatt}.
\begin{table}[t]
\begin{center}
\begin{narrowtabular}{3cm}{l|ccc}
   & Signal $a=0$  &  SM &  Background\\[0.15cm]
\hline
$ZZ$ $(4l)$ & 1.5 & 9  & 0.7 \\
$W^+ W^-$ & 5.8 & 27  & 12 \\
$W^{\pm} Z$ & 3.2 & 1.2  & 4.9 \\
$W^{\pm}W^{\pm}$ & 13 & 5.6  & 3.7
\end{narrowtabular}
\end{center}
\caption{\label{tab:WWscatt}
Event rates at the LHC ($\sqrt{s} = 14\,$TeV) with $100\, fb^{-1}$ for the scattering $VV\to VV$  
in the purely leptonic channels (with both vectors decaying to leptons). 
From ref.~\cite{Bagger:1993zf}.
Signal is defined as the difference between the total number of events predicted
for $a=0$ and that predicted in the SM ($a=1$) with $m_h = 100\,$GeV. The ``SM'' column
reports this latter number, while ``Background'' refers to the sum of 
additional reducible backgrounds.
See ref.~\cite{Bagger:1993zf} for details on the cuts and the simulation.}
\begin{center}
\begin{narrowtabular}{3cm}{l|cc}
   & Events & Significance \\[0.15cm]
\hline
 Signal $(a^2-b)=1.0$    &  7.4   &  3.7 \\ 
 Signal $(a^2-b)=0.8$    &  5.0   &  2.6 \\
 Signal $(a^2-b)= 0.5$   &  2.3   &  1.2 \\
 Background      &  1.5  &  --
\end{narrowtabular}
\end{center}
\caption{\label{tab:WWhhscatt}
Event rates at the LHC ($\sqrt{s} = 14\,$TeV) with $300\, fb^{-1}$ for the scattering $VV\to hh$ in
three-lepton final states, with the Higgs decaying to two $W$'s. From ref.~\cite{stronghh}.
The signal significance reported in the last column is computed using a Poisson statistics and 
expressed in units of standard deviations. See ref.~\cite{stronghh}
for details on the cuts and the simulation.
}
\end{table}
More recent analyses have appeared in the literature~\cite{otherWW}, also considering the
semileptonic decay modes, though none of them can yet be considered fully complete.
It is however quite clear that the study of $VV$ scattering is extremely challenging
due to the small signal cross section and the numerous SM backgrounds which can fake it.

Another smoking gun of composite Higgs models and strong symmetry breaking would be the observation
of the $VV\to hh$ scattering~\cite{Giudice:2007fh}, which in the SM has an extremely small cross section.
The importance of this channel 
comes from the fact that it is the only process giving information
on the parameter $b$, which is not constrained by the scattering $VV\to VV$ or
the precision tests and cannot be determined by measuring the single Higgs couplings.
Compared to $VV\to VV$, this process is a more ``pure'' probe of the strong dynamics,
since there is no pollution from transverse modes (the Higgs is the fourth Goldstone together
with the $W$ and $Z$ longitudinal polarizations).
An explorative analysis~\cite{stronghh} has shown that the best chances of discovery are probably in the
three-lepton final states, with the Higgs decaying to $W^+ W^-$:
$pp\to hh jj  \to 4W jj  \to l^{+}l^- l^\pm \!\not\!\! E_T \,jjjj$, see table~\ref{tab:WWhhscatt}.
A more realistic study with full detector simulation will however be needed to confirm these result
and establish the ultimate LHC sensitivity on the parameter $b$.

\acknowledgments

It is a pleasure to acknowledge the collaboration with
C.~Grojean, M.~Moretti, F.~Piccinini and R.~Rattazzi  
which led to much of the material covered by this talk.
I also~thank the organizers of IFAE 2009 for inviting me to this interesting and stimulating conference.

\end{document}